\documentstyle[preprint,epsfig,aps]{revtex}

\begin{document}
\title{The electromagnetic interaction of ultrarelativistic heavy ions}
\author{C. A. Bertulani}
\address{Department of Physics, Brookhaven National Laboratory,\\
Upton, NY 11973-5000, USA}
\maketitle

\begin{abstract}
The validity of a delta-function approximation for the
electromagnetic interaction of relativistic heavy ions is
investigated. The
production of $\ e^{+}e^{-}$ pairs, with electron capture, is
used as a test of the approximation.
\end{abstract}

\vskip 2cm

The production of $e^{+}e^{-}$ in peripheral collisions of relativistic heavy ions
has attracted a great amount of theoretical interest due
to its non-perturbative character. The calculations are hard to perform and
it is common to find substantial differences between the cross sections calculated  within
several approaches \cite{ERG00}. A good simplification of the
problem has been found by Baltz and collaborators \cite{Ba91}. 
They have shown that
if one makes a gauge transformation in the wavefunction of the form
$
\psi =\exp{\left\{ -i\chi \left( {\bf r},t\right) \right\}}\psi ^{\prime}
$,
where

\begin{equation}
\chi \left( {\bf r},t\right) =\frac{Z\alpha }{v}ln\left[ \gamma \left( z-vt\right)
+\sqrt{b^{2}+\gamma ^{2}\left( z-vt\right) ^{2}}\right]
\end{equation}
the interaction induced by the electromagnetic field of an ultrarelativistic
particle is gauge transformed to (in our units $\hbar =c=m_{e}=1 $)

\begin{equation}
V\left( \mbox{\boldmath$\rho$} ,z,t\right) =\phi \left( \mbox{ \boldmath$\rho$} ,z,t\right) \left( 1-v\widehat{%
\alpha }_{z}\right) -\phi \left(  \mbox{\boldmath$\rho$} =0,z,t\right) \left( 1-\widehat{\alpha 
}_{z}/v\right) \label{Vg}\;,
\end{equation}
where $\phi \left( \mbox{\boldmath$\rho$} ,z,t\right) $ is the Lienard-Wiechert potential at
a point ${\bf r}=\left( \mbox{\boldmath$\rho$} ,z\right) $, generated by a relativistic particle
with velocity ${\bf v}=v\hat{\bf z}$ and impact parameter ${\bf b}$,

\begin{equation}
\phi \left( \mbox{\boldmath$\rho$} ,z,t\right) =\gamma Z\alpha \left[ \left( {\bf b}-\mbox{\boldmath$\rho$} \right)
^{2}+\gamma ^{2}\left( z-vt\right) ^{2}\right] ^{-1/2}\;.\label{phi}
\end{equation}
In these expressions $\gamma =\left( 1-v^{2}\right) ^{-1/2}$ is the Lorentz
contraction factor, and $\widehat{\alpha }_{z}$ is the third of the Dirac matrices.

The second part of eq. (\ref{Vg}) acts as a regularization term of the modified
potential. It removes the divergence at $b=0$. This new potential is
very useful since the Lorentz contraction yields a
delta-function in the longitudinal variables when $\gamma \gg 1 $, and 
$b$ is not too
large. Evidently, this is a great simplification since
delta-function interactions always lead to a
considerable decrease of integration steps in perturbative as well as in
non-perturbative calculations. In \cite{Ba91} the formal derivation of the delta-function
interaction was obtained by expanding the gauge transformed potential (\ref{Vg})
into multipoles. Further manipulation of the multipole expansion and 
comparison with numerical calculations have shown that (\ref{Vg}) can be 
expressed as

\begin{equation}
V\left( \mbox{\boldmath$\rho$} ,z,t\right) =\delta \left( z-t\right) Z\alpha \left( 1-\widehat{%
\alpha }_{z}\right) \ln\frac{\left( {\bf b}-\mbox{\boldmath$\rho$} \right) ^{2}}{b^2}\, . \label{delta}
\end{equation}

We will show that this expression can be obtained in a 
simpler way. The derivation is useful to study the
validity of the delta-function approximation. In particular we will test
the approximation in a solvable problem, namely the production of $e^{+}e^{-}$ pairs in which the electron is
captured in an orbit around one of the nuclei (bound-free pairs).

Using the Bethe-integral \cite{BB88}, the potential (\ref{phi}) can be written in the form

\begin{equation}
\phi \left( \mbox{\boldmath$\rho$} ,z,t\right) =Z\alpha \, \frac{1}{2\pi ^{2}}\int d^{3}q\frac{%
e^{-i{\bf q.u}}e^{i{\bf q.r}}}{q^{2}-v^{2}q^{2}_{z}}\, , \label{bethe}
\end{equation}

where ${\bf u}={\bf b}+{\bf v}t$ and $q=\left( {\bf q}_{t},\  q_{z}\right) $. For relativistic
particles we can replace $\left( 1-v\widehat{\alpha }\right) $ and $\left( 1-%
\widehat{\alpha }/v\right) $ by $\left( 1-\widehat{\alpha }\right) $ in the
interaction (\ref{Vg}). As shown in ref. \cite{BB88} this amounts to neglect a very small ($\sim {\cal O}(1/\gamma^2)$)
piece of the longitudinal part of the interaction. However, it is important
to keep the other v factors in their respective places, as they give rise
to important combinations of $\gamma $ factors in the matrix elements. 
Moreover, the integral in (\ref{bethe}) diverges logarithmically as $v\,
\rightarrow \, 1 $. 

The exact interaction is then given by 

\begin{equation}
V\left( \mbox{\boldmath$\rho$} ,z,t\right) =Z\alpha \, \left( 1-\widehat{\alpha }_{z}\right)
\, \frac{1}{2\pi ^{2}}\int d^{3}q\ e^{-i{\bf {\bf q.u}}}\frac{\left[ e^{i{\bf q.r}}-e^{iq_{z}z}%
\right] }{q_{t}^{2}+q^{2}_{z}/\gamma ^{2}}\, , \label{full}
\end{equation}

where the denominator of the integrand in (\ref{bethe}) has been rewritten in terms of 
$\gamma $. The interaction given by eq. (\ref{delta}) is a limit of this integral when
we set $q^{2}_{z}/\gamma ^{2}=0 $. It is clear from the above equation that neglecting
this factor yields the delta-function in (\ref{delta}). However, to emphasize the
restrictions on ${\bf q}_{t} $ and $q_{z} $ let us define

\begin{equation}
\Phi \left( \mbox{\boldmath$\rho$} ,z,t\right) \equiv {V\left( \mbox{\boldmath$\rho$} ,z,t\right) \over Z\alpha \, \left( 1-%
\widehat{\alpha }_{z}\right)} =\frac{1}{\pi }\int d^{2}q_{t}\, \frac{1}{%
q^{2}_{t}}\, \exp{(-i{\bf q}_{t}.{\bf b})}\, \left[ \exp{(i{\bf q}_{t}.\mbox{\boldmath$\rho$}) }-1\right] \, \Phi
_{z}\left( {\bf q}_{t},z,t\right) \, , \label{rhoz}
\end{equation}

where

\begin{equation}
\Phi _{z}\left( {\bf q}_{t},z,t\right) =\frac{q^{2}_{t}}{2\pi }\int dq_{z}\frac{%
e^{iq_{z}\left( z-vt\right) }}{q_{t}^{2}+q^{2}_{z}/\gamma ^{2}}=\frac{\gamma
q_{t}}{2}e^{-\gamma q_{t}|z-vt|}\, .
\end{equation}

Now, using $\lim_{\Lambda \rightarrow \infty }\, \left(
\Lambda /2\right) e^{-\Lambda |x|}=\delta \left( x\right) $, we see that for 
$\gamma \rightarrow \infty $, $\Phi _{z} $ does not depend
on $q_{t} $, and assumes the form of a delta function: $\Phi _{z}\left(
z,t\right) =\delta \left( z-vt\right) $.

In this limit, we can write (\ref{rhoz})  as

\begin{equation}
\Phi \left( \mbox{\boldmath$\rho$} ,z,t\right) =\delta \left( z-t\right) 
\Phi_\rho \left( \mbox{\boldmath$\rho$}
,t\right) \, ,
\end{equation}

with

\begin{equation}
\Phi_\rho \left( \mbox{\boldmath$\rho$} ,t\right) =\frac{1}{\pi }\int d^{2}q_{t}\, \frac{1}{%
q^{2}_{t}}\, \exp{(-i{\bf q}_{t}.{\bf b})}\, \left[ \exp{(i{\bf q}_{t}.
\mbox{\boldmath$\rho$}) }-1\right] =2\int \frac{%
dq_{t}}{q_{t}}\left\{ J_{0}\left[ q_{t}|\mbox{\boldmath$\rho$} -b|\right] -J_{0}\left(
q_{t}b\right) \right\} \, ,
\end{equation}

where $J_{0} $ is the cylindrical Bessel function. The integral over each
Bessel function diverges, but their difference does not. To show
this we regularize the integrals by using

\begin{equation}
\int dx\, \frac{xJ_{0}\left( ax\right) }{x^{2}+k^{2}}=K_0\left( ak\right) \, ,
\end{equation}

where $K_{0} $ is the modified cylindrical Bessel function. Taking the limit 
$k\rightarrow 0 $, and using $K\left( ak\right) \simeq ln\left( ak\right) $,
for small values of $ak$,
we get

\begin{equation}
\Phi \left( \mbox{\boldmath$\rho$} ,t\right) =ln\frac{\left( {\bf b}-\mbox{\boldmath$\rho$} \right) ^{2}}{b^{2}}\, .
\end{equation}

This is the solution of the Coulomb potential of a unit charge in
2-dimensions. An easy way to see this is to use Gauss law for the electric
field in two dimensions. One obtains $E\simeq 1/b $, where $b $ is the
distance to the charge. Since $E=-\partial \Phi /\partial b $, the
logarithmic form of $\Phi $ is evident.

The above derivation illustrates the validity of the approximation in terms
of the transverse momentum transfer $q_{t} $. It should fail for very soft
processes, i.e., those for which $q_{t} $$\rightarrow 0. $ Also, 
it requires that $q_{z} $ is 
small compared to $\gamma $. As shown in ref. \cite{BB88}, $q_{z} $ values in
the range of one up to $\gamma$ units of the electron mass contribute appreciably to the integrals involved in
the production of free, and of bound-free,  $e^{+}e^{-} $ pairs. It is thus important to check
the validity of the approximation (\ref{delta}) in a concrete case. We will do this
for the production of bound-free pairs.
The
full calculation uses the interaction given by equation (\ref{full}). 
For comparison, a similar calculation with the term $\xi =q_{z}/\gamma $ replaced by
zero in the denominator of (\ref{full}) is equivalent to the use of the 
interaction
(\ref{delta}).

 In ref. \cite{BD00} it was shown
that the Coulomb distortion of the positron wavefunction
is an important effect in calculations 
of bound-free pair production. The right magnitude of the
differential cross sections depends on this effect. 
This is not relevant in our case, since we
are only interested in the relative change of the cross sections and
probabilities by using the exact and the delta-function
interaction, respectively. Therefore, for simplicity, we will use plane-waves for 
the positron wavefunction and
first-order perturbation theory. The energy transfer from the field to the
created pair is given by $\omega =\varepsilon +1$, where $\varepsilon $ is
the positron energy, neglecting the atomic binding energy of the
captured electron. Using the interaction in the form (\ref{full}), the amplitude for
bound-free pair production is given by

\begin{equation}
a_{fi}=i\int_{-\infty }^{\infty }dt \ e^{i\omega t}\,\left\langle \psi
_{e^{+}}\left| V\right| \psi _{e^{-}}\right\rangle =\frac{iZ\alpha }{\pi }%
\int d^{2}q_{t}\;\frac{e^{-i{\bf q}_{t}.{\bf b}}}{q_{t}^{2}+\xi ^{2}}F\left(
{\bf q}_{t}\right) \,,\label{afi}
\end{equation}
where the integral over time yields

\begin{equation}
F\left( {\bf q}_{t}\right) =\int d^{3}r\,\exp{(i\frac{\omega z}{v})}\left[
\exp{(i{\bf q}_{t}.\mbox{\boldmath$\rho$} )}-1\right] \left\langle \psi _{e^{+}}\left| \left( 1-\widehat{%
\alpha }_{z}\right) \right| \psi _{e^{-}}\right\rangle \,,
\end{equation}
and $\xi $ is now given by $\xi =\omega /\gamma v$.

Using plane waves for the positron wavefunction and a hydrogenic K-orbital
function for the electron, the above matrix element is given by

\begin{equation}
F\left( {\bf q}_{t}\right) =f\left( {\bf q}_{t}\right) \overline{{\rm v}}%
\left( 1-\widehat{\alpha }_{z}\right) u\,,
\end{equation}
where

\begin{equation}
f\left( {\bf q}_{t}\right) =8\sqrt{\pi }\left( Z\alpha \right) ^{5/2}\left\{ 
\frac{1}{\left[ 1/a_{H}^{2}+\left| {\bf Q}-{\bf p}\right| ^{2}\right] ^{2}}-%
\frac{1}{\left[ 1/a_{H}^{2}+\left| {\bf Q}_{0}-{\bf p}\right| ^{2}\right]
^{2}}\right\} \;.
\label{fq}
\end{equation}
In these equations $a_{H}=1/\alpha =5.29\times 10^{4}$ fm is the Bohr
radius, ${\rm v\;}\left( u\right) $ is the positron (electron) spinor, 
$p=\sqrt{\varepsilon ^{2}-1}$ is the
positron momentum, ${\bf Q}=\left( {\bf q}_{t},\,\omega /v\right) $, and ${\bf
Q}_{0}=\left( 0,\,\omega /v\right) $. 

Integrating the square modulus of (\ref{afi}) over ${\bf b}$ yields a delta function $%
\delta \left( {\bf q}_{t}-{\bf q}_{t}^{\prime }\right) $. Furthermore, performing the
spin averages we get for the differential cross section in terms of the
positron energy

\begin{equation}
\frac{d\sigma }{d\varepsilon d\Omega }=\frac{\left( Z\alpha \right) ^{2}}{%
2\pi ^{3}}\left( \varepsilon -1\right) p\int d^{2}q_{t}\;\frac{\left|
f\left( {\bf q}_{t}\right) \right| ^{2}}{\left( q_{t}^{2}+\xi ^2\right) ^{2}}\;. \label{dsde}
\end{equation}
The integral over the positron scattering angle can be done analytically,
as well as the remaining integral over ${\bf q}_{t}.$

To test the delta-function interaction we define the function

\begin{equation}
\Delta \left( \varepsilon \right) =\left[ \left( d\sigma /d\varepsilon
\right) _{\xi =0}-d\sigma /d\varepsilon \right] /\left( d\sigma
/d\varepsilon \right) _{\xi =0}\;, \label{D} 
\end{equation}
which does not depend on $Z$.

In figure 1 we plot the function\ $\Delta \left( \varepsilon \right) $ for
SPS, RHIC and LHC heavy ion energies. The positron energies are given in MeV
units. For SPS the above formulation applies directly, assuming that the
electron is captured in the target and neglecting the atomic screening
effects. For RHIC and LHC a transformation of (\ref{dsde}) back to the laboratory
system was performed. We notice that the delta-function interaction works very well for
positron energies of the order of MeV for SPS and RHIC and up to 100 MeV for
LHC. The approximation worsens abruptly at a certain positron
energy. The value of $\varepsilon $ \ where this occurs is a
function of $\gamma $. \ In fact, we expect that $\left( d\sigma
/d\varepsilon \right) _{\xi =0}$ starts to differ substantially
from $d\sigma
/d\varepsilon$
for $\xi$ of the order of one, i.e, for $%
\varepsilon /\gamma \simeq 1$. It is thus more appropriate to plot $%
\Delta $\ as a function of $\varepsilon /\gamma $. 
This is shown in figure 2
for the same laboratory energies as before. In this figure $\varepsilon /\gamma $ is given in
units of the electron mass. We see that all curves collapse into
approximately a single one. The differences between the results for RHIC and
for LHC are imperceptible. These results show that the calculations with the
delta-function interaction differ from the calculations \ with the exact
potential for positron energies $\varepsilon \gtrsim 0.1\gamma\ mc^{2}$. For
SPS and RHIC this implies positron energies of the order of a few MeV, and
for LHC a few hundred MeV.

In the frame of reference of the nucleus where the electron is captured, the
positrons move in the very forward direction, within an angle of the order
of $1/\gamma $ along the projectile incident direction \cite{BD00}. 
Thus, to  study the impact
parameter dependence of the production probabilities we can safely use ${\bf p%
}={\bf p}_{z}$. The differential probability for pair production is given by

\begin{equation}
\frac{dP\left( b,\varepsilon \right) }{d\varepsilon d\Omega }=\frac{\left(
Z\alpha \right) ^{2}}{8\pi ^{5}}\left( \varepsilon -1\right) p\left| \int
d^{2}q_{t}\;e^{-i{\bf q}_{t}{\bf .b}}\frac{f\left( {\bf q}_{t}\right) }{%
q_{t}^{2}+\xi ^2}\right| ^{2}\;. 
\end{equation}

Inserting the term inside brackets of eq. (\ref{fq}) in the integral above
and neglecting terms of order $1/a_H^2$, one gets the result

\begin{equation}
\frac{2\pi i}{\xi ^{2}-\eta^{2}}\left\{ \frac{1}{\xi ^{2}-\eta
^{2}}\left[ \xi K_{1}\left( \xi b\right) -\eta
K_{1}\left( \eta b\right) \right] -\frac{b}{2}K_{0}\left( \xi 
b\right) \right\} -\frac{\pi i\xi }{\eta ^{4}}K_{0}\left( \xi
b\right) \;, \label{K}
\end{equation}
where $\xi=\omega /\gamma v$ and $\eta =\omega /v-p$.

We define another function $\Delta \left( b,\varepsilon \right) $ to 
test the impact parameter dependence of the delta-function 
interaction,

\begin{equation}
\Delta \left( b,\varepsilon \right) =\left[ \left( dP/d\varepsilon d\Omega
\right) _{\xi =0}-dP/d\varepsilon d\Omega \right] /\left(
dP/d\varepsilon d\Omega \right) _{\xi =0}\;. \label{D2} 
\end{equation}
In figure 3 we plot the function\ $\Delta \left( b, \varepsilon \right) $ for
SPS, RHIC and LHC heavy ion energies, as a function of $b/\gamma $. For
comparison we use two values of the positron energy; $\varepsilon =1$ MeV,
and $\varepsilon =10$ MeV. For $\varepsilon =1$ MeV all curves
agree and one observes that the approximation is good up to impact
parameters of the order of $b\simeq0.1\gamma /\varepsilon $. This is
confirmed by looking at the curves for $\varepsilon =10$ MeV. Then
the agreement at SPS bombarding energies is not perfect, even for
the smaller impact parameters. But, for 
RHIC and LHC energies the results are basically equal. 
These results originate from the function
given in eq. (\ref{K}) which drops sharply to zero at $\xi\simeq1$,
i.e., at positron energies of the order of $b\simeq\gamma /\varepsilon 
$. This is the so-called adiabatic limit. The electromagnetic field has
photon energy components up to $1/t_{int}$, where $t_{int}$ is the
interaction time. This time equals $\left( b/\gamma \right) /v\simeq%
b/\gamma $ for relativistic collisions \cite{BB88}. 

In conclusion, the present study has shown that the
delta-function interaction yields reasonable results as long as $\omega
b/\gamma \lesssim 0.1.$ As seen in figure 3, this amounts to  
$b\lesssim 0.1\gamma /\omega $. As observed 
in ref. \cite{BB88}, the most effective impact parameters 
for this process are of the order of \ $b%
\simeq 1/m$. We also see in figures 1
and 2 that the differential cross sections $d\sigma /d\omega $ are well
described up to energies of the order of $0.1\gamma $.

For other situations, e.g.,
nuclear fragmentation due to the electromagnetic interaction in relativistic
heavy ion collisions, the most effective impact parameter is given by $b%
\simeq R$, where $R\simeq 10$ fm. We thus expect that the
delta-function interaction works well for $\varepsilon \lesssim 0.1\gamma $
MeV. Note that $\gamma$ is the Lorentz factor in the frame of reference of one
of the nuclei, i.e., $\gamma =2\gamma _{c}^{2}-1$, where $\gamma _{c}$ is
the collider Lorentz factor. Thus $\gamma$ is huge for RHIC and LHC energies,
and the approximation works well for all
energies of practical interest in nuclear fragmentation. 

The basic idea of the delta-function interaction is that the electromagnetic
field of a relativistic charge looks like a very thin 
pancake. Those processes which
do not involve too large energy transfers, will not be sensitive to the spatial
variation of the field. Then the delta-function is a good approximation. For 
typical energy transfers of the order of 10-100 MeV in 
nuclear fragmentation, the approximation works well
for $b\lesssim 0.01-0.1\gamma $ fm. To calculate total cross sections 
it is always necessary to account for those large impact parameters
at which the delta-function approximation fails. Similar conclusions have been
drawn in a recent article on projectile-electron loss in reletivistic 
collisions with atomic targets \cite{Vo00}.

\vskip 0.5 cm
{\noindent {\bf Acknowledgments}
\vskip 0.2 cm

The author is a fellow of the John Simon Guggenheim foundation. 
He is
grateful to Francois Gelis and Tony Baltz for useful discussions. 
This work has been authored under Contract N0. DE-AC02-98CH10886 with the U.S. Department of Energy. 
Partial
support from the Brazilian funding agency 
MCT/FINEP/CNPQ(PRONEX), under contract No. 41.96.0886.00, is also acknowledged.

\vskip 0.5 cm

{\noindent \bf Figure Captions}
\vskip 0.2 cm

\begin{enumerate}
\item  
Relative difference between the bound-free pair production spectrum
calculated with the delta-function approximation and with the
exact interaction, respectively (see eq. (\ref{D})). 
The comparison is done for 
SPS, RHIC and LHC heavy ion energies. The positron energies are given in MeV
units in the laboratory system. 

\item  
Same as in figure 1, but as a function of the positron energies divided
by the Lorentz gamma factor. The positron energies are given in
units of the electron rest mass.

\item  
Relative difference between the bound-free pair production probabilities
calculated with  the delta-function approximation and with the
exact interaction, respectively (see eq. (\ref{D2})). 
The comparison is done for 
SPS, RHIC and LHC heavy ion energies and for two different positron 
energies. The impact parameter divided by the Lorentz factor, 
$b/\gamma$, is given in units of the electron Compton
wavelength, $\hbar/mc$.

\end{enumerate}

\epsfig{file=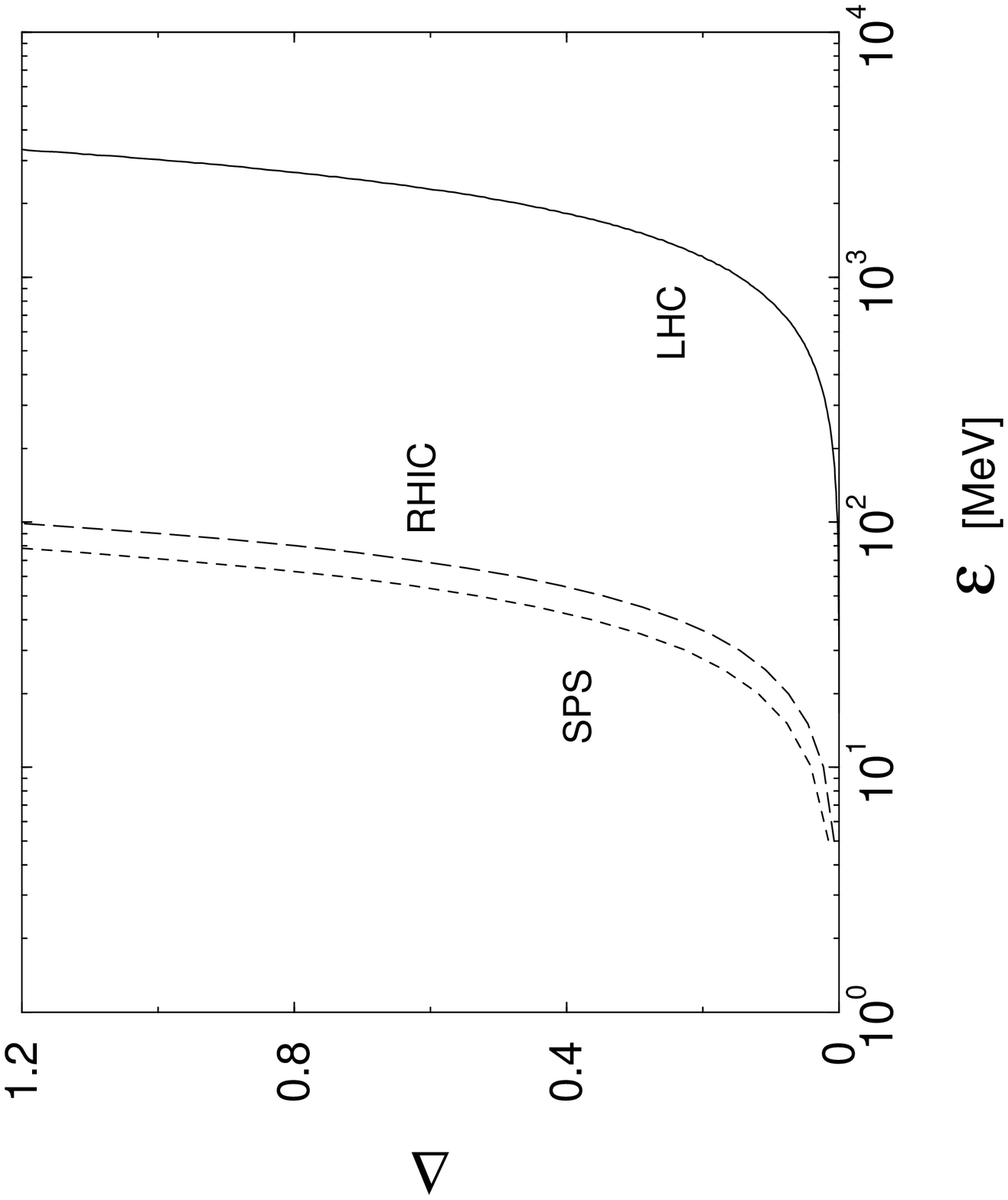,width=.75\textwidth}

\epsfig{file=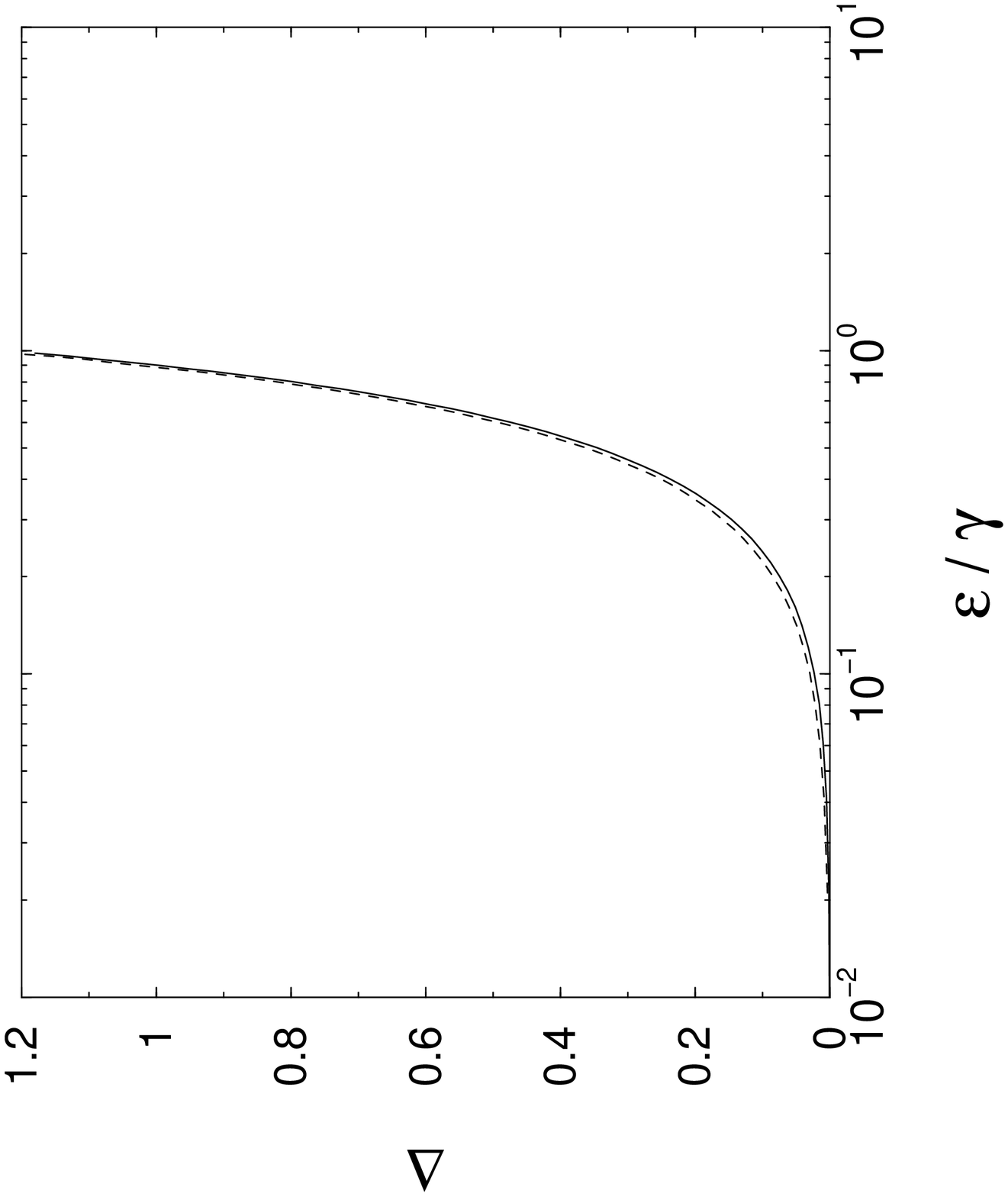,width=.75\textwidth}

\epsfig{file=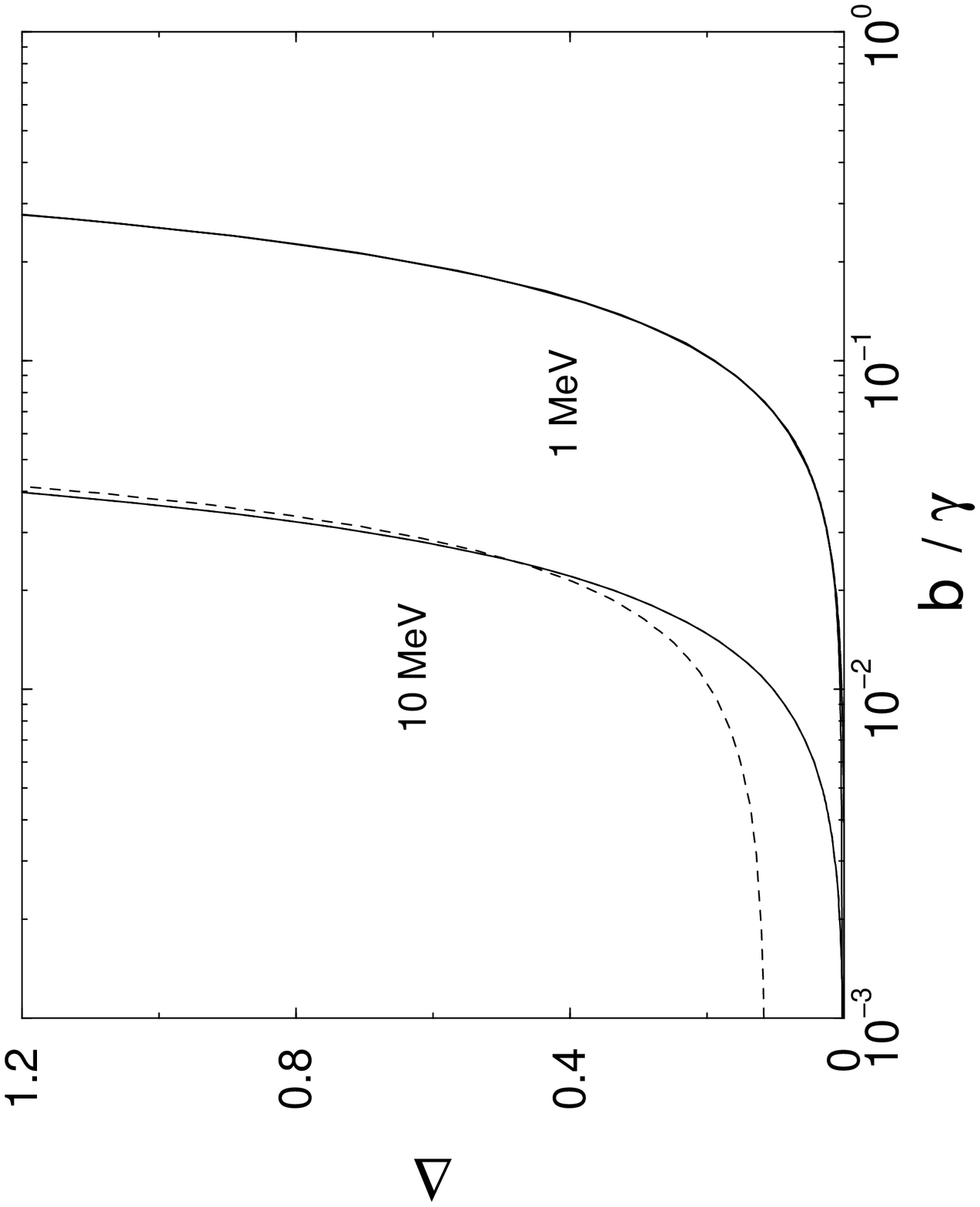,width=.75\textwidth}

\end{document}